\documentclass[%
 aip,
 amsmath,amssymb,
 reprint,%
author-numerical,%
]{revtex4-1}

\usepackage[T1]{fontenc}
\usepackage[utf8]{inputenc}

\usepackage{graphicx}
\usepackage{amssymb}
\usepackage{amsmath}
\usepackage{amsthm}

\usepackage{textgreek}

\PassOptionsToPackage{hyphens}{url}\usepackage{hyperref}

\usepackage{lineno}

\newcommand{\n}{\hat{n}}
\newcommand{\rr}{\vec{r} - \vec{r}\,'}

\begin{document}


\title{Magnetic-field modeling with surface currents: \\ Physical and computational principles of bfieldtools}



\author{Antti J Mäkinen}
\email{antti.makinen@aalto.fi}
\author{Rasmus Zetter}
\author{Joonas Iivanainen}
\author{Koos C J Zevenhoven}
\author{Lauri Parkkonen}
\author{Risto J Ilmoniemi}

\affiliation{Department of Neuroscience and Biomedical Engineering, Aalto University School of Science, FI-00076 Aalto, Finland}

\date{\today}
             

\begin{abstract}
Surface currents provide a general way to model magnetic fields in source-free volumes. To facilitate the use of surface currents in magneto-quasistatic problems, we have implemented a set of computational tools in a Python package named \texttt{bfieldtools}. In this work, we describe the physical and computational principles of this toolset. To be able to work with surface currents of arbitrary shape, we discretize the currents on triangle meshes using piecewise-linear stream functions. We apply analytical discretizations of integral equations to obtain the magnetic field and potentials associated with the discrete stream function. In addition, we describe the computation of the spherical multipole expansion and a novel surface-harmonic expansion for surface currents, both of which are useful for representing the magnetic field in source-free volumes with a small number of parameters. Last, we share examples related to magnetic shielding and surface-coil design using the presented tools.
\end{abstract}


\maketitle


\section{Introduction}
\noindent
Modeling magnetic phenomena with surface currents has various applications in physics and engineering. One large field of applications is surface-coil design, where continuous surface currents are used to design coil winding patterns. Such designs are made in plasma physics \cite{Merkel_1987, drevlak1998automated, abe2003new} magnetic resonance imaging (MRI), \cite{pissanetzky1992minimum, peeren2003stream, lemdiasov2005stream, poole2007improved, harris2013shielded, hidalgo2010theory} transcranial magnetic stimulation (TMS), \cite{koponen2017coil, cobos_sanchez_inverse_2018} magnetic particle imaging (MPI), \cite{bringout2014coil} and in zero-field magnetometry. \cite{holmes2018bi} Further applications of surface-coil design include, .e.g., field control in physics experiments \cite{afach2014dynamic, wyszynski2017active} and pickup coils of magnetic sensors. \cite{roth1990apodized, suits2003optimizing} 

The methods used in coil design are also involved in modeling eddy current patterns induced in thin conductive sheets \cite{peeren2003stream, 
zevenhoven_conductive_2014, zevenhoven_dynamical_2015} and field fluctuations due to thermal noise currents. \cite{roth_thermal_1998, uhlemann2015thermal} In addition, surface currents could be applied as equivalent models in magnetic shielding with high-permeability materials. \cite{vesanen2011spatial, Sumner1987convectional} Modeling the magnetic field in free space using equivalent current densities on the volume boundary could also have various other applications. This method can be used directly for modeling the field of uniformly magnetized bodies, \cite{blakely1996potential} or the Meissner effect in superconductors, but it could also be used as an equivalent model when interpolating magnetic-field data in e.g., in geomagnetism \cite{mendonca1994equivalent} and biomagnetism. \cite{numminen1995transformation, hanninen2001recording} Additionally, such a field model could be applied, for example, when modeling magnetic fields for interference rejection. \cite{taulu_presentation_2005}

Although surface currents are useful in modeling magnetic problems, their application has been limited because of a lack of general computational tools. Most studies have also been restricted to simple geometries.
To facilitate surface-current-based methods, we introduce a novel Python software package \texttt{bfieldtools} (available at \url{https://bfieldtools.github.io}). This package provides tools for representing currents on arbitrarily-shaped surfaces, and calculating the associated magnetic field and potentials. Further, tools for designing current patterns that generate desired magnetic fields are included. The whole software package is described in two papers. In this paper, we present physical and computational principles of the software and applications that showcase the capability of the presented tools. The accompanying paper \cite{zetter2020bfieldtools} describes the Python-based implementation in detail and provides examples of its use in different applications from the user perspective.

As in many preceding works, \cite{pissanetzky1992minimum, peeren2003stream,
abe2003new, lemdiasov2005stream, poole2007improved, zevenhoven_conductive_2014, cobos_sanchez_inverse_2018} in \texttt{bfieldtools}, we model divergence-free surface currents with scalar stream functions. We discretize these functions on a triangle mesh using piecewise-linear basis functions equivalent to piecewise-constant surface-current density.  Compared to analytical methods\cite{turner_target_1986, crozier_design_1995, brideson2002determining} that require certain symmetries for the current distributions, discretizing the stream function on a freely shaped triangle mesh allows studying currents and magnetic fields in a wide range of geometries. 

In this work, we first review the physics of the stream function. As an additional feature to previous works, we relate the stream function to harmonic potential theory. 
By introducing the magnetic scalar potential to the computational framework, analogies to other fields utilizing potential theory can be exploited, facilitating the formulation and solution of magneto-quasistatic problems.

The main objective of this work is to describe the field calculations and their discretization as they are implemented in \texttt{bfieldtools}. Based on previous studies that utilize the linear discretization of the field source, \cite{vanoosterom1983solid, demunck1992linear, ferguson1994complete, pissanetzky1992minimum} we obtain a consistent analytical discretization of the integral equations involved in the field calculations. The same principles can also be used to obtain discrete differential operators on a surface \cite{deGoes2016vector, botsch2010polygon}, which we utilize in the integral formulas. 

We have also implemented computations for series representations of the magnetic field in a source-free volume. First, we review the multipole expansion in terms of spherical harmonics, which is the conventional way of describing such a field. We adapt the multipole expansion of 3D current densities  \cite{gray1978simplified, taulu_presentation_2005, nieminen_avoiding_2011} to obtain the expansion for the field from a surface current on a mesh. In addition, we introduce a novel field representation based on expanding the stream function with the eigenfunctions of the surface-Laplacian, \cite{levy2006laplace, reuter2009} which can be seen as a generalization of the multipole expansion.

Finally, we share a few examples demonstrating the capability of these tools in coil design and magnetic shielding. More applications are described in the accompanying paper, \cite{zetter2020bfieldtools} including references to the software implementation.

\section{Stream function in quasistatic magnetism}
\subsection{Divergence-free surface currents}
\noindent A divergence-free current density $\vec j(\vec r)$ on an arbitrary surface can be expressed with a scalar stream function $\psi$ on the surface \cite{peeren2003stream,  zevenhoven_conductive_2014}
\begin{equation}
    \vec j(\vec r)  = \nabla_{\|} \psi(\vec r)  \times  \n(\vec r)\,,
    \label{eq:streamfunc_current}
\end{equation}
where $\vec{r}$ is the position on the surface, $\n$ is the unit surface normal, and $\nabla_{\|}$ the tangential gradient operator,\cite{reusken2018stream} i.e., the 3D gradient projected to a tangent plane on the surface:
${\nabla_{\|}=\nabla -\n(\n~\cdot~\nabla)}$.
  As $\nabla_{\|} \psi(\vec r)  \times  \n(\vec r)$ is perpendicular to $\nabla_{\|} \psi(\vec r)$, the streamlines of the current correspond exactly to the isocontours of $\psi({\vec{r}})$. For convenience, we define the operator $\nabla_{\|} (\cdot) \times \n$ as the \textit{rotated gradient}.

By taking a line integral of $\psi$ from $\vec{r}_0$ to $\vec{r}$ on the surface, we find that the difference in the stream function between the two ends of the path equals the flux of surface current $\vec{j}(\vec{r})$ passing the curve   \cite{peeren2003stream}
\begin{equation}
    \psi(\vec{r}) - \psi(\vec{r}_0) = \int_{\vec{r}_0}^{\vec{r}} \vec{j}(\vec{r}\,') \cdot ( d\vec{l}' \times \n')\,,
    \label{eq:stream_func_path}
\end{equation}
where $d\vec{l}'\times \n'$ is a path differential perpendicular to the direction of the path. In consequence, a line integral from a reference point $\vec{r}_0$ determines the stream function uniquely on the surface. 

From another point of view, the stream function can be interpreted as a surface density of magnetic dipoles normal to the surface\cite{peeren2003thesis, lopez2009equivalent} (see also Appendix \ref{sec:dipole_appendix}):
\begin{equation}
    \vec m(\vec r)  = \psi(\vec r) \n(\vec r)\,.
\end{equation}
This interpretation of the stream function enables analogies to dipole layers involved in, e.g., volume conductor problems and the calculation of magnetic scalar potentials for divergence-free surface currents.

As the surface current density is assumed divergence-free everywhere, the flux of current through any boundary on the surface must be zero. Applying Eq.~\eqref{eq:stream_func_path} on a boundary, we can deduce that this condition is equivalent to $\psi(\vec{r})$ being constant on the boundary. With only one boundary, the constant can be set to zero since an additional constant in $\psi(\vec{r})$ does not affect $\vec{j}(\vec{r})$. When the surface contains holes, the hole boundaries can have their own constants. These matters are further discussed in Sec.~\ref{sec:mesh_operators} when discretizing the stream function.

\subsection{Stream functions and the magnetic scalar potential}
\label{sec:bfield_surface_currents}
\noindent The magnetic field $\vec{B}$ originating from sources outside the volume of interest can be expressed as the gradient of a scalar potential $U$: $\vec{B}=-\mu_0\nabla U$. As the magnetic field is divergence-free, the magnetic scalar potential $U$ is harmonic, i.e., it satisfies Laplace's equation $\nabla^2 U = 0$. From the theory of harmonic potentials \cite{Jackson1999}, we know that $U$ can be determined uniquely in the volume (up to a constant) when either the potential or the normal derivative of the potential is specified on the boundary enclosing the volume. Thus, any external source distribution whose potential reproduces the boundary conditions of a given $U$, can be used to generate $U$ in the volume.

In particular, the boundary condition can be satisfied by the potential of a dipole density $\psi(\vec r)\n$ on the same surface. \cite{nedelec2001acoustic, hackbusch1995integral} In potential theory, this source distribution is known as a double layer, equivalent to two parallel layers of opposite charge. In magnetostatic calculations, as discussed above, such a layer of magnetic dipoles corresponds to a surface-current density $\nabla_{\|} \psi(\vec r)\times \n$. Any magnetic field within a source-free volume can thus be expressed with a stream function on the boundary of the volume. 


As the discussion above applies only to a closed surface, a stream function on a surface with openings cannot generally represent all possible  field patterns in the volume. This must be taken into account in coil designs where the current may only be placed in restricted regions as well as in field-interpolation tasks with equivalent surface currents. However, the dipole-layer analogy still applies to a stream function on an open surface: the stream function always corresponds to a discontinuity in the scalar potential \cite{zevenhoven_conductive_2014} similar to a dipole layer. \cite{nedelec2001acoustic}

\subsection{Integral equations}
\label{sec:modeling}
\noindent In the following, we layout the integral equations for calculating the quasistatic magnetic field and magnetic potentials from a stream function. The integrations are discretized in Sec.~\ref{sec:mesh_operators}. 

In source-free volumes, the magnetic field can be expressed with either a vector or scalar potential \cite{Jackson1999}
\begin{equation}
    \vec B(\vec r) = \nabla \times \vec A(\vec r) = -\mu_0 \nabla U(\vec r)\,.
\end{equation}
The vector potential of a surface current density can be written as an integral over the surface $S$, where the stream function is defined:
\begin{equation}
    \vec A(\vec r) = 
    \frac{\mu_0}{4\pi} \int_S \frac{\vec j(\vec r\,')}{|\vec r -\vec r\,'|} dS' 
    = \frac{\mu_0}{4\pi} \int \frac{\nabla_\|' \psi(\vec r\,')\times \n'}{|\vec r -\vec r\,'|} dS'\,.  
    \label{eq:vec_potential}
\end{equation}
The vector potential can be equivalently written in terms of a magnetic dipole layer $\vec{m} = \psi\n$ (see Appendix \ref{sec:dipole_appendix}), which is the more convenient form to express the magnetic scalar potential
\begin{equation}
\begin{split}
    U(\vec r) 
    &=  \frac{1}{4\pi} \int \psi(\vec{r}\,') \n' \cdot \nabla' \frac{1}{|\vec r -\vec r\,'|} dS' \,.
\end{split}
    \label{eq:potential}
\end{equation}
Similar dipole-layer potentials are used for the electric field in volume conductor problems \cite{geselowitz_bioelectric_1967}.
Finally, the Biot--Savart formula for the magnetic field is obtained as the curl of the vector potential
\begin{equation}
\begin{split}
    \vec B(\vec r) 
    &= \frac{\mu_0}{4\pi} \int (\nabla_\| \psi(\vec{r}\,')\times \n') \times \frac{\rr}{|\rr|^3} dS'\,.
    \label{eq:biot-savart}
\end{split}    
\end{equation}

In computations, it is useful to expand the stream function with a set of basis functions $\psi_k(\vec{r})$ as
\begin{equation}
    \psi(\vec{r}) = \sum_k s_k \psi_k(\vec{r})\,.
    \label{eq:streamfunc_series}
\end{equation}
The coefficients $s_k$ parametrize the stream function, enabling linear-algebraic techniques for processing it. Furthermore, the basis functions $\psi_k(\vec{r})$ can be made to satisfy possible boundary conditions so that any combination of them satisfies the same conditions. In some geometries, $\psi_k(\vec{r})$ can be chosen as, e.g., sinusoids or spherical harmonics. \cite{zevenhoven_conductive_2014, peeren2003thesis, wyszynski2017active} The rotated gradients of the basis functions provides a basis set of vector functions $\vec j_k(\vec{r}) = \nabla_{\|} \psi_k(\vec r)  \times  \n(\vec r)$ that expand the current density.

The basis function coefficients $s_k$, forming a column vector $\boldsymbol{s}$, can be used to write the inductive energy and resistive dissipation power of a surface current as quadratic forms of $\boldsymbol{s}$ \cite{peeren2003thesis, zevenhoven_conductive_2014, bringout2014coil}. The inductive energy, i.e., the energy stored in the magnetic field, can be written as $\boldsymbol{s}^{\top}\boldsymbol{M}\boldsymbol{s}/2$, where the matrix $\boldsymbol{M}$ consists of the mutual inductances of the current patterns, which can be calculated as \cite{Jackson1999}
\begin{equation}
    M_{k,l} = \int_{S} \vec j_k(\vec r) \cdot \vec A_l(\vec r) dS = \int_{S} \psi_k(\vec r)\n \cdot \vec B_l(\vec r) dS,
    \label{eq:inductive_energy}
\end{equation}
where $\vec A_l$ and $\vec B_l$ are the magnetic vector potential and the magnetic field generated by the current pattern $\vec{j}_l(\vec{r})$, respectively. 

The power dissipation due to resistive heating can be written as $\boldsymbol{s}^{\top}\boldsymbol{R}\boldsymbol{s}$, where 
\begin{equation}
    R_{k,l} = \int_{S} \vec e_k(\vec r) \cdot \vec j_l(\vec r) dS
    = \int_{S} \frac{1}{\sigma_\mathrm{s}(\vec{r})} \vec j_k(\vec r) \cdot \vec j_l(\vec r) dS
    \label{eq:dissipation_power}
\end{equation}
is the mutual resistance associated with the two current patterns. Here, $\vec{e}_k$ is the electric field associated with $\vec{j}_k$ and $\sigma_\mathrm{s}(\vec{r}) = \sigma(\vec{r}) d(\vec{r})$ is the surface conductivity defined by material conductivity $\sigma$ and surface thickness $d$. Further, assuming constant surface conductivity and using stream functions to describe $\vec{j}_k$ and $\vec{j}_l$, we get
\begin{equation}
\begin{split}
    R_{k,l}
    &= \frac{1}{\sigma_\mathrm{s}} \int_{S} [\nabla_{\|}\psi_k(\vec r)\times\n] \cdot [\nabla_{\|} \psi_l(\vec r)\times\n] dS \\
    &= \frac{1}{\sigma_\mathrm{s}} \int_{S} \nabla_{\|}\psi_k(\vec r) \cdot \nabla_{\|} \psi_l(\vec r) dS \\
    &= - \frac{1}{\sigma_\mathrm{s}} \int_{S} \psi_k(\vec r) \nabla_{\|}^2 \psi_l(\vec r) dS\,.
    \end{split}
\end{equation}
Partial integration (Gauss theorem) was used to get the last identity, where $ \nabla_{\|}^2 = \nabla_{\|}\cdot \nabla_{\|}$ is the surface Laplacian or the Laplace--Beltrami operator. \cite{reuter2009, levy2006laplace} The possible boundary terms vanish similar to derivation in Appendix \ref{sec:dipole_appendix}. The relationship between the mutual resistance and the Laplacian is utilized further in the next section.

\section{Discretization}
\label{sec:mesh_operators}
\subsection{Piecewise-linear stream function}
\label{sec:surf_current_discrete}
\noindent In \texttt{bfieldtools}, surface-current densities are represented by stream functions on triangle meshes. A triangle mesh consists of an ordered collection of vertices ${\vec{r}_1,..., \vec{r}_V}$, forming a point cloud in a 3D space, and of a set of triangular faces $\Delta_f$, each defined by a triplet $(i,j,k)$ of vertex indices.

\begin{figure}[t]
    \centering
    \includegraphics[width=\columnwidth]{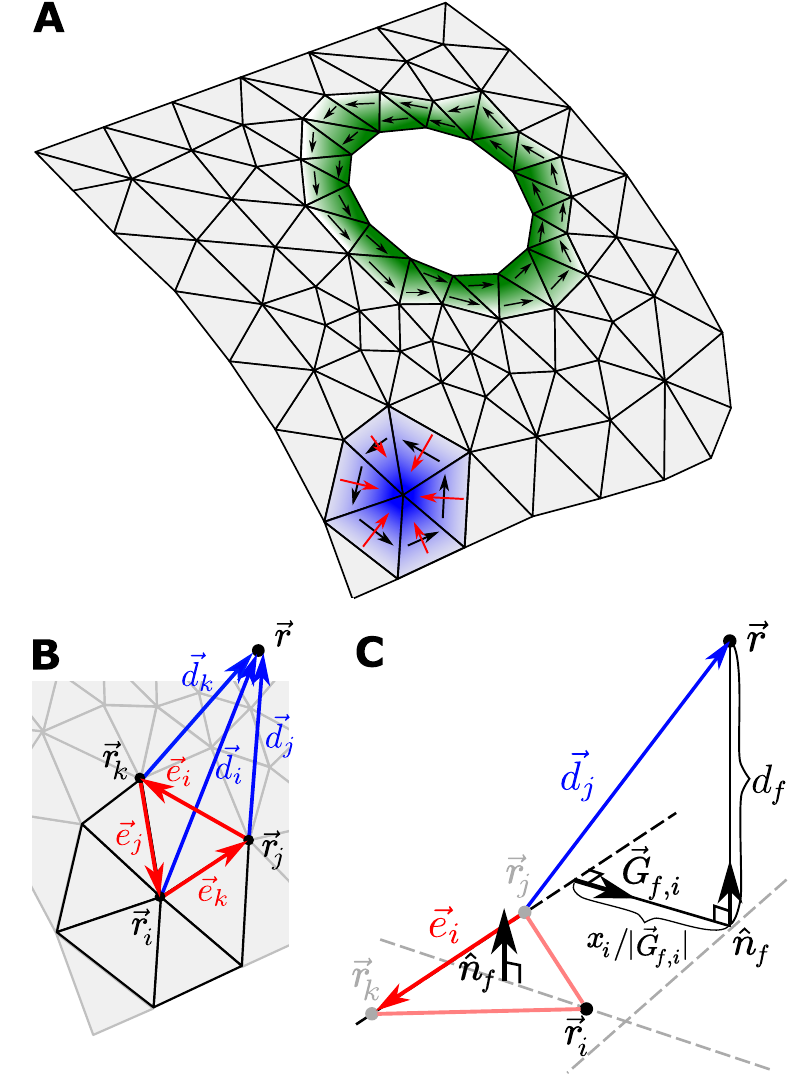}   
    \caption{\textbf{A}: In the lower left corner of the triangle mesh, the blue color indicates a hat function with an increasing function value towards the center vertex. The red arrows represent the gradient of the function. The black arrows correspond to the rotated gradient constituting an eddy current around the vertex. The green color around the hole indicates a basis function for the hole, constructed to satisfy the constant boundary condition along the hole boundary. \textbf{B}: Euclidean vectors used in the differential operators and integral formulas. \textbf{C}: Geometry for the signed distance functions $x_i = \vec{G}_{f,i}\cdot\vec{d}_j$ and $d_f=\n_{f}\cdot\vec{d}_j$ used in the analytical integral formulas.} 
    \label{fig:hatfunc}
\end{figure}

We discretize the integral and differential equations described in the previous section by approximating the stream function as linear on each face of the triangle mesh. Such piecewise-linear functions can be conveniently expressed as in Eq.~\eqref{eq:streamfunc_series} by choosing the basis functions $\psi_k(\vec{r})$ to be so-called hat functions $h_i(\vec{r})$, where the index $i$ corresponds to the $i$th vertex of the mesh. The hat function $h_i(\vec{r})$ has the value one at vertex $i$ and zero at all other vertices. Within triangles, the value is interpolated linearly (see Fig. \ref{fig:hatfunc}A). 
As in Eq.~\eqref{eq:streamfunc_series}, the stream function can be written as a sum of the basis functions
\begin{equation}
    \psi(\vec r) = \sum_i s_i h_i(\vec r)\,,
    \label{eq:psi_hat}
\end{equation}
where $s_i$, the weight for the vertex $i$, is equal to the current circulating around the vertex on the neighbouring triangles. We obtain current-density basis functions by taking the rotated gradient of the hat function $\vec{j}_i(\vec{r}) = \nabla_{\|} h_i(\vec r)\times \n\,$, which corresponds to an eddy current circulating around vertex $i$ as illustrated using black arrows in Fig.~\ref{fig:hatfunc}A. 

In each neighbouring triangle, the gradient and rotated gradient of a hat function are constant    
and can be expressed using the local geometry \cite{deGoes2016vector, botsch2010polygon} as
\begin{align}
    &\nabla_{\|} h_i(\vec{r}) =  \n_f \times \frac{\vec{e}_i}{2A_f}\,,
    \label{eq:gradh} \\
    &\nabla_{\|} h_i(\vec{r}) \times \n_f =  \frac{\vec{e}_i}{2A_f}\,.
    \label{eq:gradr}
\end{align}
where $A_f$ is the area of the neighbouring triangle $f$, $\n_f$ the triangle normal and $\vec{e}_i$ the edge opposing the vertex $i$ in the triangle.

The constant condition on the outer mesh boundary can be implemented by setting the boundary-vertex values to zero. However, each hole boundary can float at an arbitrary value. To satisfy the constant boundary condition on the hole boundaries, we construct a combined basis function for each hole boundary $C_k$ as
\begin{equation}
    h_{C_k}(\vec r) = \sum_{i\in C_k} h_i(\vec{r})\,.
    \label{eq:hole_function}
\end{equation}
These functions are constant along the hole boundaries and can be conceptualized as a current flowing around the hole within the triangles neighbouring the hole vertices (Fig.~\ref{fig:hatfunc}A). The stream function can now be expressed as
\begin{equation}
    \psi(\vec{r}) = \sum_i s_i h_i(\vec{r}) + \sum_k s_k h_{C_k}(\vec r)\,,
    \label{eq:stream_discrete_holes}
\end{equation}
where the first part sums over the inner vertices of the mesh and the latter sums over the holes of the mesh.

With this vertex-wise discretization of the stream function, we can represent physical quantities using operators acting on the vertex values $s_i$. Stacking the weights $s_i$ into a column vector $\boldsymbol{s}$, linear operators (e.g. the surface-Laplacian) acting on $\psi$ discretize to matrices that can be used in linear mappings $\boldsymbol{s}\mapsto\boldsymbol{As}$ or in quadratic forms $\boldsymbol{s}\mapsto\boldsymbol{s}^{\top}\boldsymbol{A}\boldsymbol{s}$. Additionally, fields originating from the discretized current can be expressed as $\vec{B}(\vec r) = \sum_n \vec{B}_n(\vec r)\,s_n = \vec{\boldsymbol{B}}(\vec r)^\top \boldsymbol{s}$, where $\vec{\boldsymbol{B}}(\vec r)$ is a column vector of the magnetic field contributions at $\vec{r}$ from each vertex in the mesh. 

\subsection{Differential operators}
\label{sec:diff_operators}
\noindent With the hat-function discretization, the gradient of any scalar function can be calculated on the faces of the mesh from the neighbouring vertex values. \cite{botsch2010polygon, deGoes2016vector} As this calculation is linear with respect to the vertex values, we define a discrete gradient operator $\boldsymbol{\vec{G}}$ as a map from scalar values at the vertices ($i$) to Euclidean vectors at the faces ($f$). Since there are only three non-zero hat functions on each triangle, the result of this operation can expressed as
\begin{equation}
    (\boldsymbol{\vec{G}}\boldsymbol{s})_f = \sum_l \vec{G}_{f, l} s_l = \vec{G}_{f, i} s_i + \vec{G}_{f, j} s_j + \vec{G}_{f, k}s_k\,,
\end{equation}
where $i$, $j$, and $k$ are the vertices of $\Delta_f$. One element of the operator is obtained directly from the gradient of the basis function Eq.~\eqref{eq:gradh} as
\begin{equation}
    \vec{G}_{f, i} =  \begin{cases} \n_f \times \frac{\vec{e}_i}{2A_f},\, &i\in\Delta_f \\
    0, &i\notin \Delta_f  \,.
    \end{cases}
    \label{eq:grad_operator}
\end{equation}
The elements of the rotated-gradient operator are defined as 
\begin{equation}
    \vec{G}^\perp_{f, i} = \vec{G}_{f, i} \times \n_f\,.
    \label{eq:gradr_operator}
\end{equation}

\begin{figure}[t]
    \centering
    \includegraphics{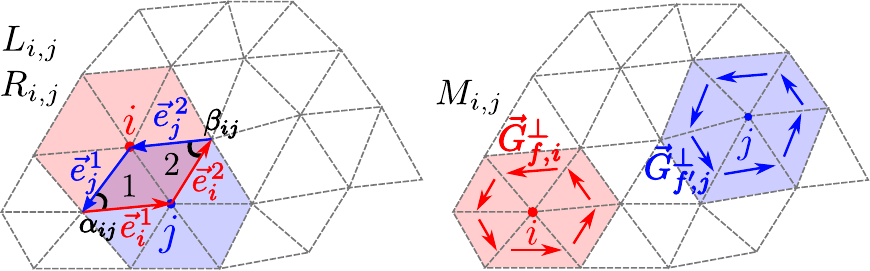}
    \caption{Geometric quantities in the discretization of inductance, resistance and Laplacian operators. Laplacian and resistance matrices are sparse: only the vertices whose neighbouring triangles overlap contribute to to the matrix elements. The inductance matrix is dense with elements describing the coupling between the currents circulating the vertices.}
    \label{fig:operators}
\end{figure}

Using the hat functions to discretize the surface-Laplacian operator leads to the so-called cotan formula derived and applied in the context of partial differential equations as well as in geometry and graphics processing. \cite{macneal1949solution, pinkall1993computing, jacobson2013algorithms, crane2013geodesics} As second derivatives are ill-defined for hat functions (zero on the faces, infinite on vertices and edges), the discrete Laplacian operator $\boldsymbol{L}$ is understood as the weak (integrated) form of the surface Laplacian:
\begin{equation}
\begin{split}
    L_{i,j} &= - \int \nabla_\|h_i(\vec{r}) \cdot\nabla_\| h_j(\vec{r}) dS \\ &= -\sum_f
    (\vec{G}_{f,i} \cdot \vec{G}_{f,j}) A_f,
    \label{eq:laplace_weak_form}
\end{split}
\end{equation}
Using Eq.~\eqref{eq:grad_operator}, the non-zero off-diagonal elements of $\boldsymbol{L}$ can be expressed as
\begin{equation}
\begin{split}
    L_{i,j} &= -\frac{1}{2}\left(\frac{\vec{e}_i^1 \cdot \vec{e}_j^1}{2A_1} + \frac{\vec{e}_i^{\,2} \cdot \vec{e}_j^{\,2}}{2 A_2} \right)
    \\ &= -\frac{1}{2}[\cot(\alpha_{ij}) + \cot(\beta_{ij})]
\end{split}
    \label{eq:laplace_operator}
\end{equation}
where $i$ and $j$ correspond to two neighbouring vertices, angles $\alpha_{ij}$ and $\beta_{ij}$ are the angles opposing the edge connecting the vertices, and indices 1 and 2 correspond to the two triangles that share the edge, as illustrated in Fig.~\ref{fig:operators}. Since constant functions belong to the null space of the Laplacian, the diagonal elements can be obtained as
\begin{equation}
    L_{i,i}  = - \sum_{j \neq i} L_{i,j}\,.
\end{equation}

When the surface has boundaries (outer edges or holes), the Laplacian has to be modified. Elements that correspond to the zero-valued boundary can be left out of the matrix. Using Eqs.~\eqref{eq:hole_function} and \eqref{eq:stream_discrete_holes}, it can be deduced that the elements that correspond to the basis function of a hole boundary can be obtained by summing the rows and columns associated with the vertices on the boundary.

\subsection{Analytical integrals}
\label{sec:integrals}
\begin{figure}[t]
    \centering
    \includegraphics[width=\columnwidth]{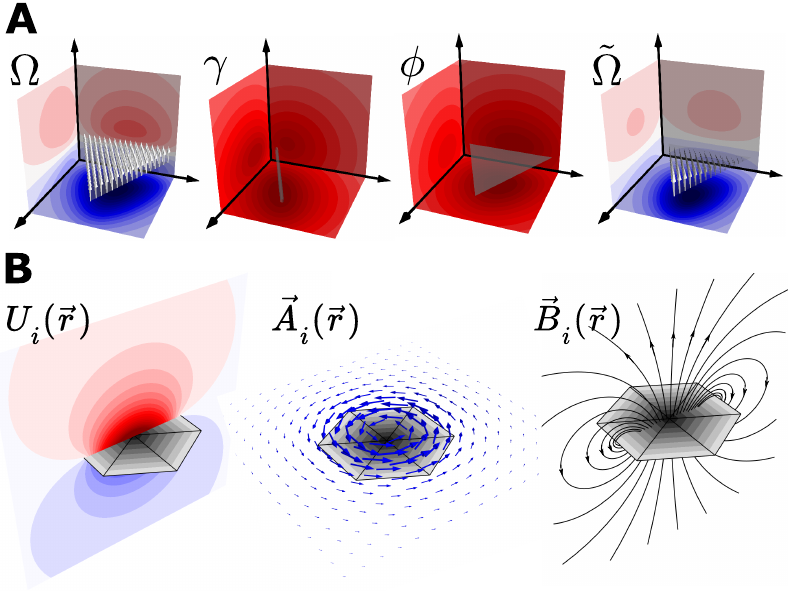}
    \caption{\textbf{A}: Building blocks of the field operators visualized as source configurations and their respective potentials. The integral $\Omega$ is the potential of a uniform dipole density (white arrows) on a triangle, $\gamma$ is the potential of a line charge, $\phi$ is the potential of a uniform charge density on a triangle, and $\hat{\Omega}$ is the potential of a linearly varying dipole density on a triangle. \textbf{B}: The magnetic field $\vec{B}_i$, scalar potential $U_i$ and vector potential $\vec{A}_i$ calculated for a single stream-function element. The stream function (a hat function) is represented by the gray color on the six triangles. The magnetic field (black lines) $\vec{B}_i$ and the scalar potential $U_i$ (red-blue colors) are visualized on the same vertical plane and the vector potential (blue arrows) is shown on a horizontal plane above the stream function element. Due to the analytical formulas, the computed fields are well-behaving in the vicinity of the mesh.}
    \label{fig:integrals}
\end{figure}

\noindent The analytical integral formulas introduced in this section are the basic building blocks of \texttt{bfieldtools} mesh operators for the magnetic field and magnetic potentials (Sec.~\ref{sec:modeling}). As the integrals needed to compute the mesh operators involve singular quantities, the analytical formulas behave more smoothly in the proximity of the source mesh compared to numerical quadratures. These formulas have been derived in the literature related to boundary-element methods in bioelectromagnetism \cite{vanoosterom1983solid, demunck1992linear, ferguson1994complete} and in antenna modeling \cite{rao1979simple, wilton1984potential, graglia1993numerical}.
To introduce concepts and to unify notation, we review the analytical formulas, which can also be seen as potentials of simple charge or dipole configurations visualized in Fig.~\ref{fig:integrals}A.

The first building block, used in all the field calculations, is the solid angle subtended at $\vec r$ by triangle $\Delta_{f}$ consisting of vertices $i$, $j$, and $k$: \cite{vanoosterom1983solid}
\begin{equation}
    \Omega_{f}(\vec r) = - \int_{\Delta_{f}}\frac{\rr}{|\rr|^3} \cdot d \vec S\,'= - 2\,\mathrm{atan2}(N,D)\,,
    \label{eq:solid_angle}
\end{equation}
where  $N = \vec d_{i}\times \vec d_{j} \cdot \vec d_{k}$ is the numerator and $D = |\vec d_i||\vec d_{j}||\vec d_{k}| + \sum_{l=(i,j,k)} |\vec d_l|(\vec d_{l+1}\cdot\vec d_{l-1})$ the denominator for the two-argument inverse tangent function $\mathrm{atan2}$ defined as in most standard programming languages. Here, $\vec d_i = \vec r - \vec r_i$ is a vector pointing from the vertex $i$ to the evaluation point $\vec{r}$ (see Fig.~\ref{fig:hatfunc}B). The magnetic scalar potential of unit (magnetic) dipole density on a triangle can be computed using the solid angle as $-\Omega_f/(4\pi)$.

The second building block is the potential of a line charge. The potential of a unit line charge on edge $\vec e_i$  (Fig.~\ref{fig:hatfunc}B) can be obtained as \cite{demunck1992linear}
\begin{equation}
    \gamma_i(\vec r) = \frac{1}{|\vec e_i|}\int_{\vec r_j}^{\vec r_k}\frac{1}{|\vec r -\vec r\,'|} dl' 
    = \frac{-1}{|\vec e_i|}\ln \frac{|\vec d_{j}||\vec e_{i}| + \vec d_{j}\cdot\vec e_{i}}{|\vec d_{k}||\vec e_{i}| + \vec d_{k}\cdot\vec e_{i}}\,,
    \label{eq:gamma}
\end{equation}
where $\vec r_j$ and $\vec r_k$ are the two ends of the edge $\vec{e}_i$.

With the two integrals above, we can express the potential of a unit charge density on a triangle $\Delta_{f}$: \cite{ferguson1994complete}
\begin{equation}
\begin{split}
    \phi_{f}(\vec r) &= \int_{\Delta_{f}}\frac{1}{|\vec r -\vec r\,'|} dS\,' \\
    =& d_f(\vec{r})\Omega_{f}(\vec{r}) + \sum_{l=i,j,k} 2 A_f x_l(\vec{r}) \gamma_l(\vec{r})\,.
    \label{eq:phi}
\end{split}
\end{equation}
Here, $d_f=\n_f \cdot \vec{d}_{l+1}$ is the signed distance from the triangle plane along the plane normal and $x_l = \vec{G}_{f,l} \cdot \vec{d}_{l+1}$ is the normalized signed distance from the line extended from the edge $\vec{e}_l$ along $\vec{G}_{f,l}$ such that $x_l(\vec{r}_l)=1$. The geometry related to these distances can be found in Fig.~\ref{fig:hatfunc}C. 

Finally, we present the potential of a linearly varying dipolar density $h_i(\vec r)$ on a triangle $\Delta_{f}$: \cite{demunck1992linear}
\begin{equation}
\begin{split}
    \tilde{\Omega}_{f,i}(\vec r) &= \int_{\Delta_{f}} h_i(\vec r\,')\frac{\vec r -\vec r\,'}{|\vec r -\vec r\,'|^3}\cdot d \vec S\,'\\
    =&  -x_i(\vec{r}) \Omega_{f}(\vec{r}) + \sum_{l=i,j,k} c_{i,l}  d_f(\vec{r}) \gamma_l(\vec{r}) \,,
    \label{eq:omega_tilde}
\end{split}
\end{equation}
where $c_{i,l} = \vec{e_i} \cdot \vec{e_l} /(2A_f)$. As the notation in Eqs.~\eqref{eq:phi} and \eqref{eq:omega_tilde} differ from the literature, we provide alternative, concise derivations of the formulas in Appendix \ref{sec:Bfield_appendix} using the notation of this work.

\subsection{Magnetic field and magnetic potentials} \label{sec:field_and_pots}
\noindent We express the magnetic field and potentials using mesh operators such that, e.g., the magnetic field  at $\vec{r}$ is $\vec{B}(\vec r) = \sum_i \vec{B}_i(\vec r)\,s_i = \vec{\boldsymbol{B}}(\vec r)^\top \boldsymbol{s}$, where the sum is taken over the vertices of the mesh and $\vec{B}_i(\vec{r})$ is the magnetic field corresponding to hat function $h_i(\vec{r})$. When a set of field evaluation points $\{\vec{r}_j\}$ is given, the operators can be expressed as coupling matrices, whose elements equal the coupling between the hat-function currents and the field components at the evaluation points.


The magnetic field $\vec{B}_i(\vec{r})$ of a constant current density in a triangle is derived in \ref{sec:Bfield_appendix} and, using that result, the magnetic field due to a single hat-function current becomes
\begin{equation}
\begin{split}
    \vec{B}_i(\vec{r}) 
    &= \frac{\mu_0}{4\pi}\sum_{f \in\mathcal{N}_i} \int_{\Delta_f} \vec{G}^\perp_{f,i} \times \frac{\vec{r}- \vec{r}\,'}{|\vec{r}- \vec{r}\,'|^3} dS' \\
        &= \frac{\mu_0}{4\pi}\sum_{f \in\mathcal{N}_i}\left(\Omega_{f}(\vec{r}) \vec{G}_{f,i} - \sum_{l=i,j,k} c_{i,l} \gamma_l(\vec{r}) \n_f \right)\,,
    \end{split}
\end{equation}
where $\mathcal{N}_i$ denotes the set of triangles neighbouring vertex $i$ as shown in Fig.~\ref{fig:operators}. A corresponding formula expressed in local coordinates of a triangle has been derived by Pissanetzky.~\cite{pissanetzky1992minimum}

The vector and scalar potentials for the current of a hat stream function $h_i$ can be obtained in a straightforward manner using the integrals in Sec.~\ref{sec:integrals}. The vector potential [Eq.~\eqref{eq:vec_potential}] can be expressed using the discrete rotated-gradient [Eq.~ \eqref{eq:gradr_operator}] and the integral $\phi_f(\vec{r})$: [Eq.~\eqref{eq:phi}] \cite{pissanetzky1992minimum, koponen2017coil}
\begin{equation}
\begin{split}
    \vec{A}_i(\vec{r}) 
    &= \frac{\mu_0}{4\pi}\sum_{f\in\mathcal{N}_i} \int_{\Delta_f} \frac{\vec{G}^\perp_{f, i}}{|\vec{r}- \vec{r}\,'|} dS' 
    =\frac{\mu_0}{4\pi}\sum_{f \in\mathcal{N}_i} \vec{G}^\perp_{f,i}\phi_f(\vec{r})\,.
        \end{split}
    \label{eq:vec_pot_discrete}
\end{equation}
The scalar potential [Eq.~\eqref{eq:potential}] of $h_i$ involves only the potentials of linearly varying dipole densities $\tilde{\Omega}_{f,i}(\vec{r})$ [Eq.~\eqref{eq:omega_tilde}]
\begin{equation}
\begin{split}
    U_i(\vec{r}) 
    &=  \frac{1}{4\pi}\sum_{f\in\mathcal{N}_i} \int_{\Delta_f} h_i(\vec{r}\,')\n \cdot \frac{\vec{r}- \vec{r}\,'}{|\vec{r}- \vec{r}\,'|^3} dS' \\
    &=\frac{1}{4\pi} \sum_{f \in\mathcal{N}_i} \tilde{\Omega}_{f,i}(\vec{r})\,.
        \end{split}
\end{equation}
The magnetic field and potentials due to a single hat-function current are illustrated in Fig.~\ref{fig:integrals}B.

\subsection{Mutual inductance and resistance}
\label{sec:energy_operators}
\noindent 
The mutual inductance between two hat-function currents (Fig.~\ref{fig:operators}) can be calculated using Eq.~\eqref{eq:inductive_energy} as
    \begin{equation}
    \begin{split}
    M_{i,j} &=  \frac{\mu_0}{4\pi} \sum_{f\in\mathcal{N}_i}\sum_{f' \in\mathcal{N}_j} \int_{\Delta_f} \int_{\Delta_{f'}} \frac{\vec{G}^\perp_{f, i}\cdot\vec{G}^\perp_{f', j}}{|\vec{r}- \vec{r}\,'|} dS dS' \\
    &=  \frac{\mu_0}{4\pi} \sum_{f\in\mathcal{N}_i}\sum_{f'\in\mathcal{N}_j}  \vec{G}^\perp_{f,i}\cdot\vec{G}^\perp_{f', j} \sum_q w_q \phi_f(\vec{r}_q)\,,
    \end{split}
\end{equation}
where the second integral is calculated using quadrature points $\vec{r}_q$ with weights $w_q$ as calculated by Koponen \cite{koponen2017coil}. We have implemented this approach in \texttt{bfieldtools}, as it naturally handles the singularity in the double integral when $\Delta_f = \Delta_{f'}$. Alternatively, the singularity can be handled with an analytical formula for the self element. \cite{eibert1995calculation} 

In this basis, the mutual inductance operator $\boldsymbol{M}$ can also be interpreted as a mapping from the discretized stream function $\boldsymbol{s}$ to the magnetic flux (integrated normal component) at the mesh vertices. This can be seen from the second identity in Eq.~\eqref{eq:inductive_energy}: by replacing $\vec{m}_k$ with the dipole density $\n h_k$, the matrix element $M_{k,l}$ corresponds to the normal magnetic field of current $l$ integrated over the hat function of vertex $k$.

For mutual resistance we also have to model the surface conductivity $\sigma_\mathrm{s}$. Assuming piecewise-constant surface conductivity on the triangles, and using  Eq.~\eqref{eq:dissipation_power}, we obtain the mutual resistance operator as
\begin{equation}
\begin{split}
    R_{i,j} &= \int \frac{1}{\sigma(\vec{r}) d} \nabla_\|h_i(\vec{r}) \cdot\nabla_\| h_j(\vec{r}) dS 
     \\ &=  \sum_f 
     \frac{\vec{G}_{f,i} \cdot \vec{G}_{f,j}}{\sigma_f d} A_f
     = \frac{1}{2d}\left(\frac{\vec{e}_i^1 \cdot \vec{e}_j^1}{2A_1\sigma_1} + \frac{\vec{e}_i^{\,2} \cdot \vec{e}_j^{\,2}}{2 A_2 \sigma_2} \right)\,,
\end{split}
\end{equation}
where $\sigma_1$ and $\sigma_2$ are the conductivities in triangles 1 and 2 neighbouring the edge from vertex $i$ to vertex $j$ (see Fig.~\ref{fig:operators}). When $\sigma$ is constant over the surface, the resistance operator is proportional to the discrete surface Laplacian
    $R_{i,j} = -\frac{1}{\sigma_\mathrm{s}}L_{i,j}\,$.
Several studies \cite{lemdiasov2005stream, poole2007improved, lopez2009equivalent} use mutual resistance in this form
although the relation to the discrete Laplacian has not been noted.

\section{Magnetic field representations with source expansions}
\label{sec:magnetic}
\noindent The magnetic field in free space can be expanded as a series of components, each of which can be interpreted to correspond to a certain type of a source-current pattern. A common series used for the static magnetic field is the spherical multipole expansion. This expansion can represent a spatially smoothly varying magnetic field with a few parameters, which can be helpful, e.g., when designing coils that generate these types of fields. \cite{nieminen_avoiding_2011, Xia2017theory, wyszynski2017active} 

A disadvantage of the multipole expansion is, however, that the series can diverge in regions where the actual field is well-behaving. For more general purposes, we introduce a representation of magnetic fields based on a stream function, which can be viewed as an equivalent source of the field. We expand the stream function on a surface as a series of functions that we call surface harmonics. The magnetic field patterns of the surface harmonics then yield a representation of the field similar to the multipole expansion.

\begin{figure}
    \centering
    \includegraphics[width=\columnwidth]{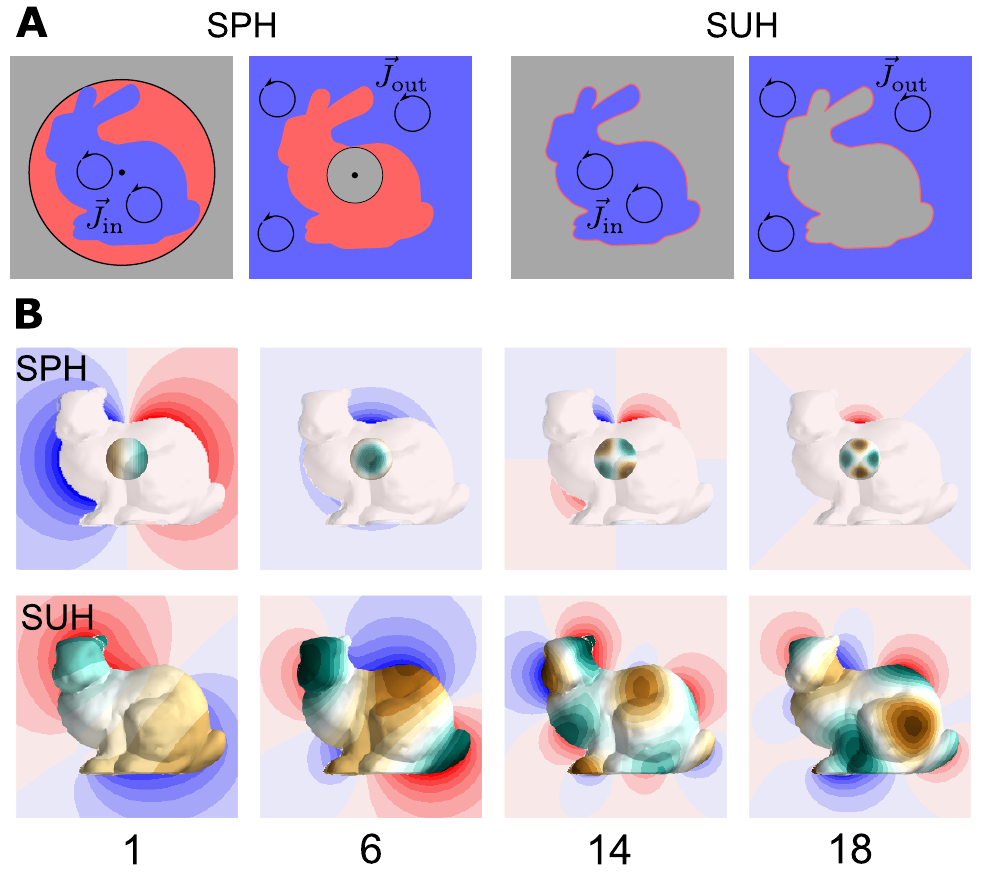}
    \caption{\textbf{A}: Convergence regions of the spherical-harmonic multipole expansion (SPH) and the surface harmonic expansion (SUH) shown in gray. Blue regions depict the volumes of magnetic sources and red regions the volumes where the expansions do not converge. \textbf{B}: The magnetic scalar potential corresponding to different components of the spherical harmonic (SPH) and surface harmonic (SUH) inner-source expansions. The scalar potential is depicted by the red--blue colors on the vertical plane and the field source is illustrated by the green--brown color. In both expansion, the first, fifth, 11$^\mathrm{th}$, and 16$^\mathrm{th}$ component of the series are shown. The bunny surface mesh was decimated from the original (http://graphics.stanford.edu/data/3Dscanrep/\#bunny) and repaired for holes.}
    \label{fig:expansions}
\end{figure}

\subsection{Multipole expansion with spherical harmonics}
\label{sec:multipole}
\noindent The general solution of Laplace's equation in spherical coordinates ($\vec{r}, \theta, \varphi$) is \cite{Jackson1999}
\begin{equation}
    U(r, \theta,\varphi) = \sum_{l=0}^\infty \sum_{m=-l}^l (\alpha_{lm}r^{-l-1} + \beta_{lm}r^l) Y_{lm}(\theta,\varphi)\,,
    \label{eq:spherical_harmonic_potential}
\end{equation}
where $\alpha_{lm}$ and $\beta_{lm}$ are the multipole coefficients, $Y_{lm}(\theta,\varphi)$ are spherical harmonic functions with degree $l$ and order $m$ ($|m| \leq l$). The $\alpha_{lm}$ terms involve powers of the inverse distance, representing sources close to origin, whereas the $\beta_{lm}$ terms represent far-away sources. In \texttt{bfieldtools}, we use the real spherical harmonics, \cite{plattner2014spatiospectral} which are orthonormal with respect to integration over the full solid angle, i.e, $\int_\Omega Y_{lm} Y_{l'm'} d\Omega = \delta_{ll'}\delta_{m m'}\,$. 

We obtain the expansion for the magnetic field by taking the gradient of the scalar potential:
\begin{equation}
\begin{split}
    \vec B(\vec r) &= -\mu_0 \nabla U(\vec r)  \\
    &= -\mu_0\sum_{l=0}^\infty \sum_{m=-l}^l [\alpha_{lm} \nabla(r^{-l-1} Y_{lm}(\theta,\varphi)) \\
    &\quad\qquad\qquad\quad+ \beta_{lm}  \nabla(r^l  Y_{lm}(\theta,\varphi))]  \\
    &= -\mu_0\sum_{l=0}^\infty \sum_{m=-l}^l [\alpha_{lm}r^{-l-2} \sqrt{(l+1)(2l+1)}\vec V_{lm}(\theta,\varphi)\\ 
    &\qquad\qquad\qquad\;+ \beta_{lm}r^{l-1} \sqrt{l(2l+1)} \vec W_{lm}(\theta,\varphi)]\,,
\end{split}
\end{equation}
where $\vec V_{lm} = (-(l+1)Y_{lm}\hat{r} + \nabla_1 Y_{lm})/\sqrt{(l+1)(2l+1)}$ and $\vec W_{lm}= (l Y_{lm}\hat{r} + \nabla_1 Y_{lm})/\sqrt{l(2l+1)}$ are vector spherical harmonics. \cite{hill_theory_1954, taulu_presentation_2005} Here, $\nabla_1$ is the angular part of the gradient on a unit sphere. \cite{plattner2014spatiospectral} Both the set of vector spherical harmonics $\vec V_{lm}$, $\vec W_{lm}$, and the set of tangential vector spherical harmonics $\vec X_{lm} = -\hat{r} \times \nabla_1 Y_{lm} /\sqrt{l(l+1)}$ are also orthonormal with respect to an inner product $\int_\Omega \vec{f}_{lm}\cdot\vec{g}_{l'm'}d\Omega\,$.

Because the multipole expansion of $\vec{B}$ is linear with respect to the coefficients, we can express it using linear operators $\boldsymbol{\vec{B}}_\alpha$ and $\boldsymbol{\vec{B}}_\beta$ as
\begin{equation}
    \vec{B}(\vec{r}) = \boldsymbol{\vec{B}}_\alpha(\vec{r})^\top\boldsymbol{\alpha} + \boldsymbol{\vec{B}}_\beta(\vec{r})^\top\boldsymbol{\beta}\,,
\end{equation}
where the expansions coefficients, truncated at a certain degree $l$, are stacked in the column vectors $\boldsymbol{\alpha}$ and $\boldsymbol{\beta}$. When the magnetic field is evaluated at a specific set of evaluation points, the linear operators above can be expressed as coupling matrices that map the given multipole coefficients to field values at the evaluation points.

The coefficients $\alpha_{lm}$ and $\beta_{lm}$ can be calculated directly from any surface-current distribution $\vec{j}(\vec{r})$ with help of the tangential vector spherical harmonics $\vec X_{lm}$ as \cite{gray1978simplified, nieminen_avoiding_2011}
\begin{align}
    \alpha_{lm} &= \frac{\sqrt{l(l+1)}}{(l+1)(2l+1)}\int  (r')^{l} \vec X_{lm}(\vec{r\,'}) \cdot \vec j(\vec r') dS' 
    \label{eq:alpha} \\
    \beta_{lm} &= -\frac{\sqrt{l(l+1)}}{l(2l+1)} \int  (r\,')^{-l-1} \vec X_{lm}(\vec{r}\,') \cdot \vec j(\vec r\,') dS'\,.
\end{align}
Using these equations together with the stream function in Eq.~\eqref{eq:streamfunc_current} and its discretization in Eq.~\eqref{eq:psi_hat}, we define mesh operators (matrices) $\boldsymbol{C}_\alpha$ and $\boldsymbol{C}_\beta$ that map the stream-function values to the spherical harmonic coefficients
\begin{align}
\boldsymbol{\alpha} &= \boldsymbol{C}_\alpha\boldsymbol{s} \\ \boldsymbol{\beta} &= \boldsymbol{C}_\beta\boldsymbol{s}\,. 
\end{align}


The convergence of the multipole series may be analyzed by inserting, for example, the inner multipole coefficients of Eq.~\eqref{eq:alpha} into Eq.~\eqref{eq:spherical_harmonic_potential} which yields terms involving factors $(r'/r)^{(l-1)}$. If there are any sources with radius $r'$ greater than the radius of the field point $r$, the factors $(r'/r)^{(l-1)}$ approach infinity with growing $l$ and the series fails to converge.
A similar analysis can be made for the outer sources. These analyses result in convergence regions for the expansions shown in Fig.~\ref{fig:expansions}A. As the figures show, the choice of the origin is crucial for the convergence, but it cannot be chosen in such a way that the series would converge everywhere in the volume where the scalar potential is defined.

\begin{figure}
    \centering
    \includegraphics[width=\columnwidth]{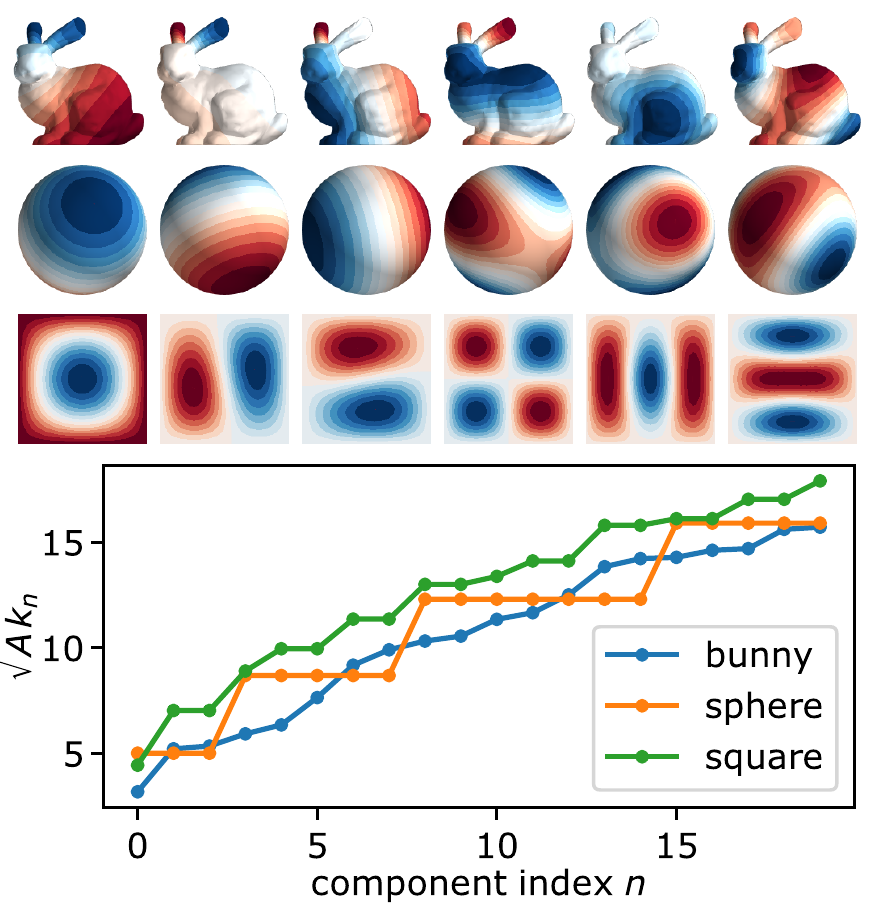}
    \caption{Surface-harmonic basis functions, i.e. eigenfunctions of the surface Laplacian, for three surfaces and their eigenvalue spectra normalized with the square root of the surface area ($\sqrt{A}$). Shown are the first six basis functions for each surface obtained by numerically solving Eq.~\eqref{eq:suh_eigen}.}
    \label{fig:suh_bases}
\end{figure}


\subsection{Surface-harmonic expansion}
\label{sec:suh_expansion}
\noindent With the tools presented in this work, we can generate a field expansion that converges at all points where the magnetic scalar potential is defined [Fig.~\ref{fig:expansions}]. The expansion is based on the fact that any potential satisfying the Laplace equation can be written in terms of an equivalent stream function on a boundary of the domain as described in Sec.~\ref{sec:bfield_surface_currents}. Expanding the stream function as a linear combination of basis functions with increasing order of spatial detail yields a field representation similar to the spherical multipole expansion.

We base the expansion on the eigenfunctions of the (negative) surface Laplacian. These eigenfunctions generalize a sinusoidal function basis, such as the spherical harmonics basis, to an arbitrary surface (see Fig. \ref{fig:suh_bases}). Generally, these functions are characterized by the eigenvalue equation \cite{levy2006laplace, reuter2009}  
\begin{equation}
   - \nabla_{\|}^2 v_n(\vec{r}) =  k_n^2 v_n(\vec{r})\,,
\end{equation}
where the eigenvalue $k_n^2$ corresponds to the squared spatial frequency of the $n^\mathrm{th}$ eigenfunction $v_n$. The higher the order $n$ is, the higher the spatial frequency and the more zero crossings $v_n(\vec{r})$ has (Fig.~\ref{fig:suh_bases}). In relation to spherical harmonics, we call these functions \textit{surface harmonics} (SUHs). In geometry processing, they are also known as manifold harmonics. \cite{vallet2008spectral}

For practical computations, we discretize the functions as $v_n(\vec{r}) = \sum_i V_{i,n} h_i(\vec{r})$, which leads to a discrete (generalized) eigenvalue equation \cite{reuter2009}
\begin{equation}
    - \boldsymbol{L} \boldsymbol{v}_n = k_n^2 \boldsymbol{N} \boldsymbol{v}_n\,,
    \label{eq:suh_eigen}
\end{equation}
where $\boldsymbol{L}$ is the Laplace operator in Eq.~\eqref{eq:laplace_operator} and $\boldsymbol{N}$ is a mass matrix taking into account the overlap of hat functions, and $\boldsymbol{v}_n$ correspond to columns of the matrix $\mathbf{V}$. As both $\boldsymbol{L}$ and $\boldsymbol{N}$ are sparse matrices, the vertex coefficient vectors $\boldsymbol{v}_n$ can be solved efficiently with sparse solvers. The resulting eigenfunctions $v_n(\vec{r})$ are orthonormal with respect to integration over the surface, which can be expressed in the discrete form as $\boldsymbol{v}_n^\top\boldsymbol{N}\boldsymbol{v}_m = \delta_{n,m}$.

Substituting the surface-harmonics representation of a stream function $\psi(\vec r) = \sum_n a_n v_n(\vec{r})$ to Eq.~\eqref{eq:potential}, we can write the magnetic scalar potential as
\begin{equation}
\begin{split}
U(\vec{r})  &= \frac{1}{4\pi} \int \psi(\vec{r}\,') \n' \cdot \nabla' \frac{1}{|\vec r -\vec r\,'|} dS' \\
    &= \sum_i a_i \frac{1}{4\pi} \int v_n(\vec r\,') \n' \cdot \nabla' \frac{1}{|\vec r -\vec r\,'|} dS' \\
    &= \sum_{i,n} a_i V_{i,n} \frac{1}{4\pi} \int h_i(\vec r\,') \n' \cdot \nabla' \frac{1}{|\vec r -\vec r\,'|} dS' \\
    &= \sum_{i,n} a_i V_{i,n} U_i(\vec{r}) = \boldsymbol{U}(\vec r)^\top\boldsymbol{V}\boldsymbol{a}\,.
\end{split}
\end{equation}
Similarly, the SUH coefficients $\boldsymbol{a}$ can be mapped to the magnetic field as
\begin{equation}
    \vec{B}(\vec{r}) = \boldsymbol{\vec B}(\vec r)^\top\boldsymbol{V}\boldsymbol{a}\,.
\end{equation}
Examples of the scalar potentials of the basis functions $v_n$ (SUH) are displayed in Fig.~\ref{fig:expansions}B with a comparison to the multipole expansion (SPH). Compared to the multipole potentials, in the SUH expansion, the potentials are distributed more uniformly around the corresponding surface. 

The SUH expansion is not restricted to closed surfaces, but can be applied for stream functions on surfaces with boundaries and any number of holes. 
Such bases can be used for surface-coil design to decrease the degrees of freedom when optimizing surface currents, as demonstrated in the accompanying paper.  \cite{zetter2020bfieldtools}


Instead of orthogonal stream functions, one may desire orthogonality in their magnetic fields. In that case, the basis functions can be solved from an discrete eigenvalue equation similar to Eq.~\eqref{eq:suh_eigen} by replacing the mass matrix $\boldsymbol{N}$ with the inductance matrix $\boldsymbol{M}$. To enable physical interpretations, $\boldsymbol{L}$ can be replaced by $\boldsymbol{R}=\boldsymbol{L}/\sigma_\mathrm{s}$ to get the following eigenvalue equation
\begin{equation}
    \boldsymbol{R}\boldsymbol{s}_n = \frac{1}{\tau_n} \boldsymbol{M}\boldsymbol{s}_n.
\end{equation}
This equation is related to the independent modes of eddy currents on the conducting surface; $\tau_n$ is the time constant of the $n$th mode. These modes can be used, for example, to calculate the time dynamics of eddy-current induced fields \cite{zevenhoven_conductive_2014} or uncoupled current patterns for thermal noise calculations. \cite{roth_thermal_1998, iivanainen2020noise} It should be noted, however, that $\boldsymbol{M}$ is now a dense matrix, whereas $\boldsymbol{N}$ was very sparse, disabling the use of sparse eigensolvers and increasing computation time when building the matrix.


\section{Coil design and shielding}
\noindent 
In this section, we give examples that utilize the developed tools. As the design of surface coils using distributed currents is probably the most prominent application of these tools, we start by giving a brief overview of the coil-design method. In surface-coil design using triangle meshes, \cite{lemdiasov2005stream, poole2007improved, cobos_sanchez_inverse_2018} the coil current is expressed with a discretized stream function $\boldsymbol{s}$ on the mesh similarly as in our tools. The stream-function $\boldsymbol{s}$ is optimized by minimizing a cost function while taking into account given constraints for, e.g., the field shape. Finally, the coil wires are placed on the isocontours of the stream function to approximate the continuous current density.
    
Typically, a quadratic form of $\boldsymbol{s}$ such as the magnetic energy $\boldsymbol{s}^\top\boldsymbol{M}\boldsymbol{s}$ or the dissipated power $\boldsymbol{s}^\top\boldsymbol{R}\boldsymbol{s}$ is used as a cost function.
Constraints for the field pattern can be formulated using the mesh operator $\boldsymbol{\vec B}(\vec{r})$ and they can be incorporated in quadratic programming as demonstrated in the accompanying paper. \cite{zetter2020bfieldtools}

Here, in the next examples, we take a more theoretical approach to surface-coil design and, in particular, to the design of self-shielded coils. We also share an example of a calculation related to magnetic shielding using the tools described in this work.

\subsection{Perfect shielding by surface currents on a closed surface }
\label{sec:surface_currents}
\begin{figure}[t]
    \centering
    \includegraphics[width=0.48\textwidth]{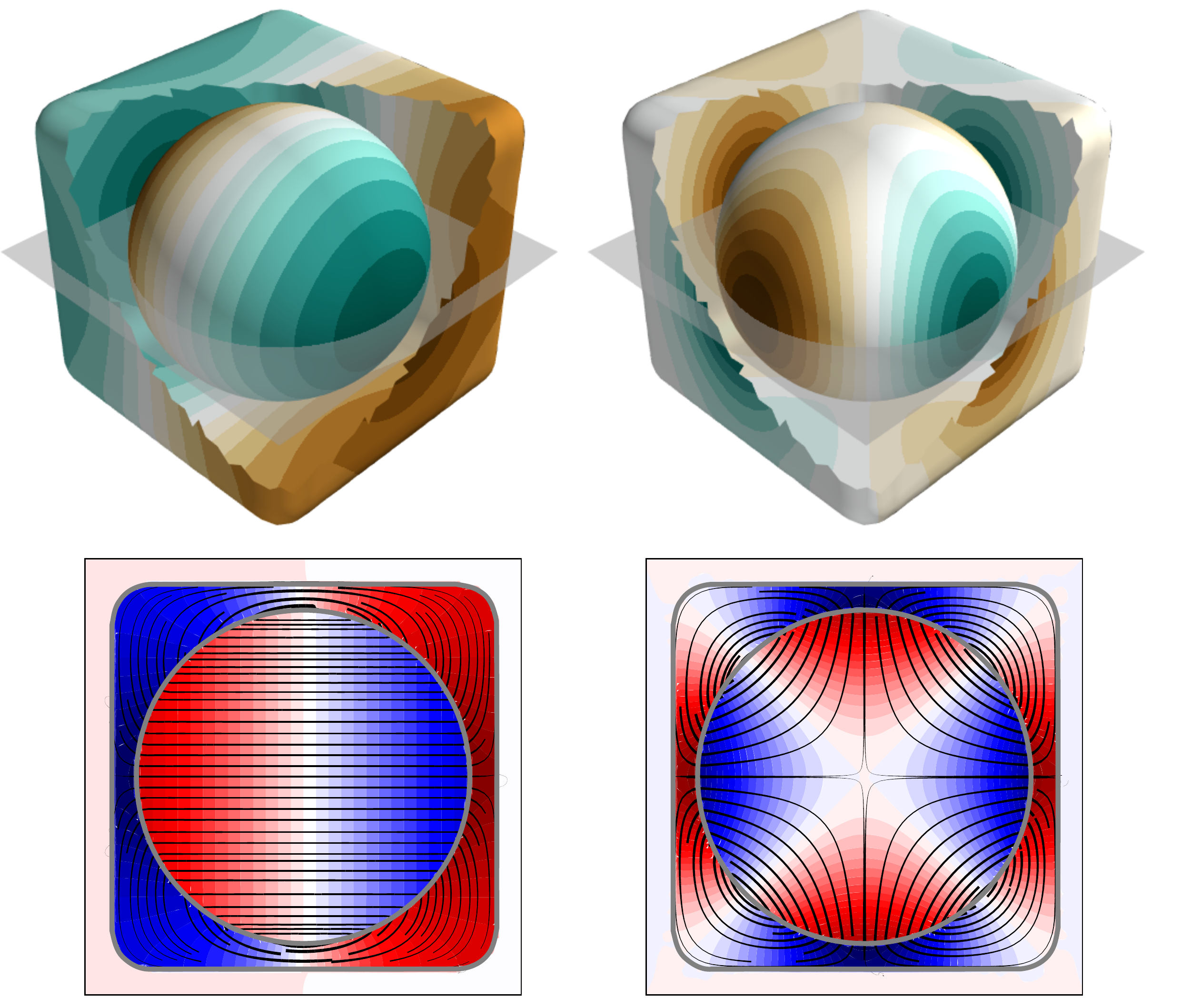}
    \caption{Two examples of surface currents on two surfaces, generating a desired magnetic field confined in a closed volume. \textbf{Top:} Stream functions of two surface-current configurations with a primary current on the sphere and a shielding current on the rounded cube. The currents are designed so that together they create a homogeneous field (left) and a first-order gradient field (right). \textbf{Bottom:} the magnetic field lines (black) and the corresponding magnetic scalar potential are plotted on the horizontal plane shown in the 3D plots on top.}
    \label{fig:shielded_field}
\end{figure}




\begin{figure*}[t]
    \centering
    \includegraphics[width=\textwidth]{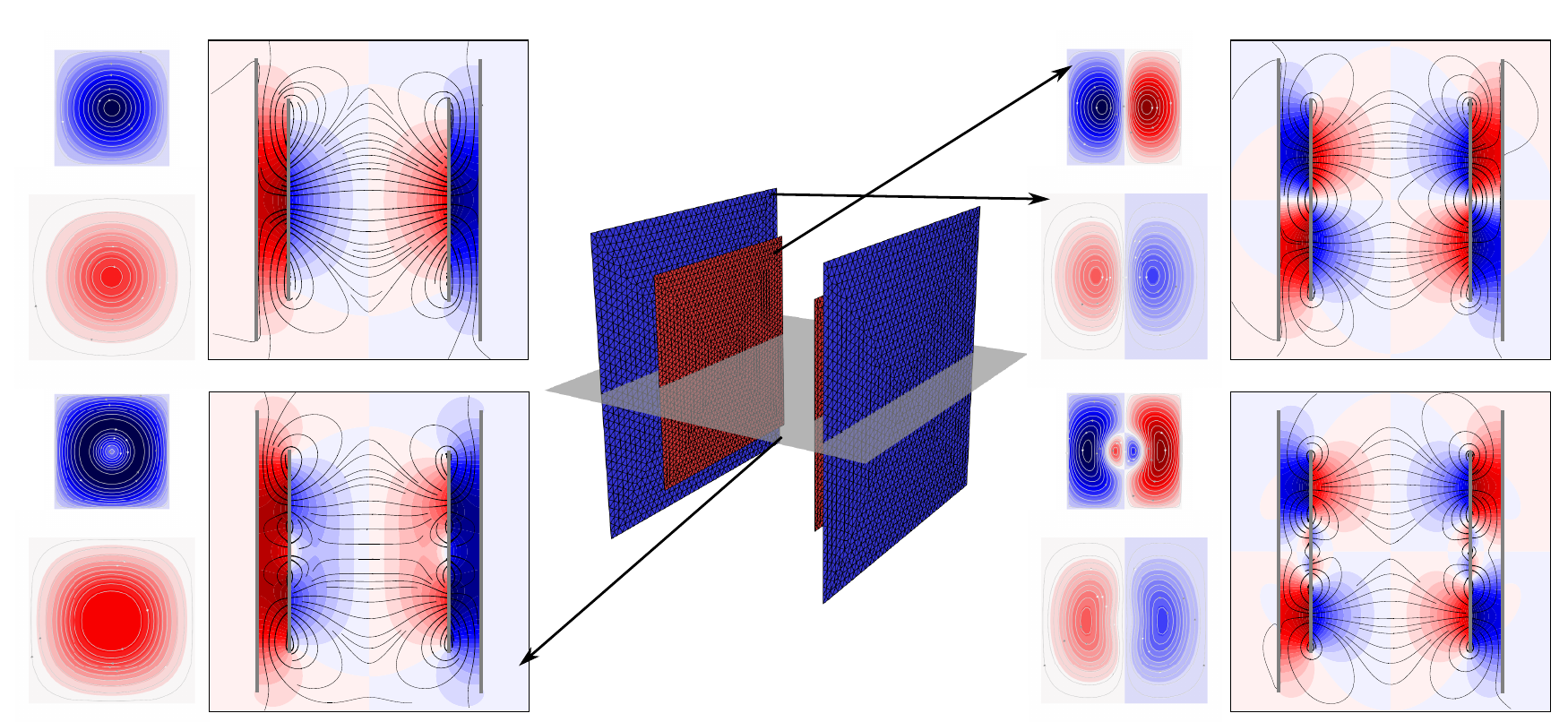}
    \caption{Shielded magnetic fields designed for bi-planar current-domains obtained by minimizing the quadratic objective in Eq.~\eqref{eq:objective_s1}. \textbf{Top left}: Homogeneous field with minimum inductance weighting (small $\lambda$). \textbf{Top right}: Homogeneous field with minimum multipole residual weighting (large $\lambda$). \textbf{Bottom left}: Gradient field with minimum inductance weighting. \textbf{Bottom right}: Gradient field with minimum multipole residual weighting. In each case, the magnetic field is visualized using the stream lines and a contour plot of the scalar potential on the horizontal plane shown in the middle. The associated stream functions on the primary and shielding surfaces are shown left from the field plots.}
    \label{fig:shielded_biplanar}
\end{figure*}

\noindent
When designing coils for target magnetic fields, it is often also desired to control the field outside the volume of interest, e.g., to shield the external environment from the field of the primary current. For such a situation, a shielding current outside the primary surface can be designed. 

Let us consider a closed surface, inside which a desired field pattern is to be designed and a second (outer) surface, the exterior of which is to be shielded from the field. To derive a set of equations with a unique solution, we again discretize the surfaces using triangle meshes. The shielded field pattern can be obtained by designing suitable stream functions $\boldsymbol{s}_1$ and $\boldsymbol{s}_2$ on the two surfaces so that their combined field satisfies desired boundary conditions. Based on the discussion in Sec.~\ref{sec:energy_operators}, we can write the boundary conditions for the normal component of the field at the surfaces using stacked mutual-inductance matrices $\boldsymbol{M}_{ij}$ as
\begin{equation}
    \boldsymbol{M}\boldsymbol{s} =
    \begin{bmatrix}
    \boldsymbol{M}_{11} & \boldsymbol{M}_{12} \\
    \boldsymbol{M}_{21} & \ \boldsymbol{M}_{22}
    \end{bmatrix}
    \begin{bmatrix} \boldsymbol{s}_1 \\ \boldsymbol{s}_2
    \end{bmatrix} = 
     \begin{bmatrix} \boldsymbol{b}_\mathrm{n} \\ \boldsymbol{0}
    \end{bmatrix}\,,
    \label{eq:bc_shielded}
\end{equation}
where the first row corresponds to the desired magnetic field  $\boldsymbol{b}_\mathrm{n}$ at the inner surface and the second row corresponds to the zero condition for the outer surface. Because $\boldsymbol{s}^{\top}\boldsymbol{M}\boldsymbol{s}$ is the total magnetic-field energy of the system, $\boldsymbol{M}$ is a positive semi-definite matrix. As the only zero eigenvalue is the one corresponding to a constant stream function (zero current), the system can be solved by inverting $\boldsymbol{M}$ deflated for the constant vector.

Two examples demonstrating the perfect shielding obtained by solving Eq.~\eqref{eq:bc_shielded} are shown Fig.~\ref{fig:shielded_field}. The shielding by the outer surface corresponds to the situation where the exterior volume would be a superconductor that expels all fields so that no magnetic field crosses the outer surface. This is also analogous to an electrical volume-conductor problem where the exterior volume is insulating, confining the 3D current density.

When the outer current surface contains current-free regions in it, perfect cancellation of the primary field is generally not possible due to a lack of degrees of freedom in the current-pattern design. Next, we demonstrate a method to optimize the primary and shielding currents applicable also to an open surface geometry.


\subsection{Self-shielded currents with an open geometry}
\label{sec:coil_design}
\noindent We now apply the tools presented in this work for the design of self-shielded coils in a more realistic bi-planar geometry. Consider a primary coil with stream function $\boldsymbol{s}_1$ and a shielding coil with stream function $\boldsymbol{s}_2$. Demonstrated by Harris and coworkers\cite{harris2013shielded}, a well-performing shielding coil can be designed by minimizing the magnetic-field energy with respect to $\boldsymbol{s}_2$. Using $\boldsymbol{s}_1$ and $\boldsymbol{s}_2$, the field energy can then be expressed as
\begin{equation}
\begin{split}
    E_{\mathrm{M}} &= \frac{1}{2}
    \begin{bmatrix} \boldsymbol{s}_1^\top & \boldsymbol{s}_2^\top
    \end{bmatrix}
    \begin{bmatrix}
    \boldsymbol{M}_{11} & \boldsymbol{M}_{12} \\
    \boldsymbol{M}_{21} & \ \boldsymbol{M}_{22}
    \end{bmatrix}
    \begin{bmatrix} \boldsymbol{s}_1 \\ \boldsymbol{s}_2
    \end{bmatrix}\\
    &= \frac{1}{2} \boldsymbol{s}_1^\top\boldsymbol{M}_{11}\boldsymbol{s}_1 + 
    \boldsymbol{s}_1^\top\boldsymbol{M}_{21}\boldsymbol{s}_2 +
     \frac{1}{2} \boldsymbol{s}_2^\top\boldsymbol{M}_{22}\boldsymbol{s}_2 \,.
\end{split}
\end{equation}
The minimum (for given $\boldsymbol{s}_1$) can be found by equating the gradient of $E_{\mathrm{M}}$ with respect to $\boldsymbol{s}_2$ to zero, which yields a set of linear equations:
\begin{equation}
    \boldsymbol{M}_{21}\boldsymbol{s}_1 +
    \boldsymbol{M}_{22}\boldsymbol{s}_2 = 0\,.
    \label{eq:zero_flux}
\end{equation}
Based on Eq.~\eqref{eq:bc_shielded}, we can now explain why this method works: the equation can be interpreted as a condition that the normal self-field of the shielding current $\boldsymbol{M}_{22}\boldsymbol{s}_2$ exactly cancels the normal field component generated by the primary current $\boldsymbol{M}_{21}\boldsymbol{s}_1$ at the shielding surface. By solving Eq.~\eqref{eq:zero_flux} for $\boldsymbol{s}_2$, we can rewrite the field energy as $E_{\mathrm{M}} = \frac{1}{2} \boldsymbol{s}_1^\top(\boldsymbol{M}_{11} - \boldsymbol{M}_{12}\boldsymbol{M}_{22}^{-1}\boldsymbol{M}_{21})\boldsymbol{s}_1 =  \frac{1}{2} \boldsymbol{s}_1^\top\boldsymbol{\tilde{M}}_{11}\boldsymbol{s}_1$. 

We will now optimize the primary current $\boldsymbol{s}_1$ for minimal energy with an additional constraint. Instead of a hard equality constraint for the desired field, we modify the cost function with a term that penalizes for the residual in desired multipole moments $\boldsymbol{\beta}$: 
\begin{equation}
 E(\boldsymbol{s}_1) = \frac{1}{2}\boldsymbol{s}_1^\top \tilde{\boldsymbol{M}}_{11}\boldsymbol{s}_1 + \lambda\|\boldsymbol{\beta}-\boldsymbol{C}_\beta \boldsymbol{s}_1\|^2,
 \label{eq:objective_s1}
\end{equation}
where $\lambda$ is a trade-off parameter between a perfect multipole fit and a minimal field energy. The coupling matrix $\boldsymbol{C}_\beta = \boldsymbol{C}_{\beta,1} + \boldsymbol{C}_{\beta,2}\boldsymbol{M}_{22}^{-1}\boldsymbol{M}_{21}$ is obtained using the constraint in Eq.~\eqref{eq:zero_flux}; $\boldsymbol{C}_{\beta,i}$ are matrices that map $\boldsymbol{s}_i$ to the multipole moments.
The solution for $\boldsymbol{s}_1$ that minimizes the cost function can again be found by equating the gradient $\nabla_{\boldsymbol{s_1}} E(\boldsymbol{s}_1)$ to zero:
\begin{equation}
    \boldsymbol{s}_1 = (\boldsymbol{C}_\beta^\top \boldsymbol{C}_\beta + \boldsymbol{\tilde{M}}_{11}/\lambda) ^{-1}\boldsymbol{C}_\beta\boldsymbol{\beta}\,.
\end{equation}
 
Examples of shielded configurations generated using the method above are shown in Fig.~\ref{fig:shielded_biplanar}. Compared to the fields in Fig.~\ref{fig:shielded_field}, which are solved for a closed geometry, these fields leak in the directions where the shielding surface is missing. Thus, the placement of the shielding surfaces is crucial for the self-shielding performance.

\subsection{Modeling a high-permeability magnetic shield}
\noindent Magnetic measurements are usually shielded from the low-frequency fluctuations of the outside magnetic environment with soft ferromagnetic materials such as $\mu$-metal. The purpose of these materials is to guide the external magnetic field to create a magnetic void inside the shield. A downside is that the shield also distorts the fields generated inside the shield. When the relative permeability of the shield is very large, the effect of the shield can be approximated by a shield with infinite relative permeability. This leads to a boundary condition stating that the magnetic field must be normal to the inner surface of the shield or, equivalently, the inner surface has to be at equipotential in terms of the magnetic scalar potential. \cite{Jackson1999}

Let us consider a primary potential $\boldsymbol{U}_\mathrm{p}(\vec{r})^\top\boldsymbol{s}_\mathrm{p}$ generated by a surface current with (discretized) stream function $\boldsymbol{s}_\mathrm{p}$ inside the magnetic shield. We can satisfy the equipotential condition on the shield by placing a suitable equivalent surface current $\boldsymbol{s}_\mathrm{eq}$ on the shield surface. In other words, we require that $\boldsymbol{U}_\mathrm{p}^\top(\vec{r})\boldsymbol{s}_\mathrm{p}+ \boldsymbol{U}_\mathrm{eq}(\vec{r})^\top\boldsymbol{s}_\mathrm{eq} = 0$ holds on the shield. To solve for $\boldsymbol{s}_\mathrm{eq}$, we apply the condition at collocation points on the shield mesh. Because $\boldsymbol{U}_\mathrm{eq}(\vec{r})\boldsymbol{s}_\mathrm{eq}$ is discontinuous across the shield, we apply the condition at collocation points slightly inwards from the shield surface as $\vec{r_j}  - \epsilon\n_j$, where $\vec{r}_j$ are the vertex positions of the shield mesh, $\epsilon$ is a small number compared to the mesh resolution and $\vec{n}_j$ are the vertex normal vectors. After solving the resulting set of linear equations for $\boldsymbol{s}_\mathrm{eq}$, we can estimate the effect of the shield by  $\boldsymbol{U}_\mathrm{eq}(\vec{r})^\top\boldsymbol{s}_\mathrm{eq}$ at all points inside the shield.

\begin{figure}[!t]
    \includegraphics{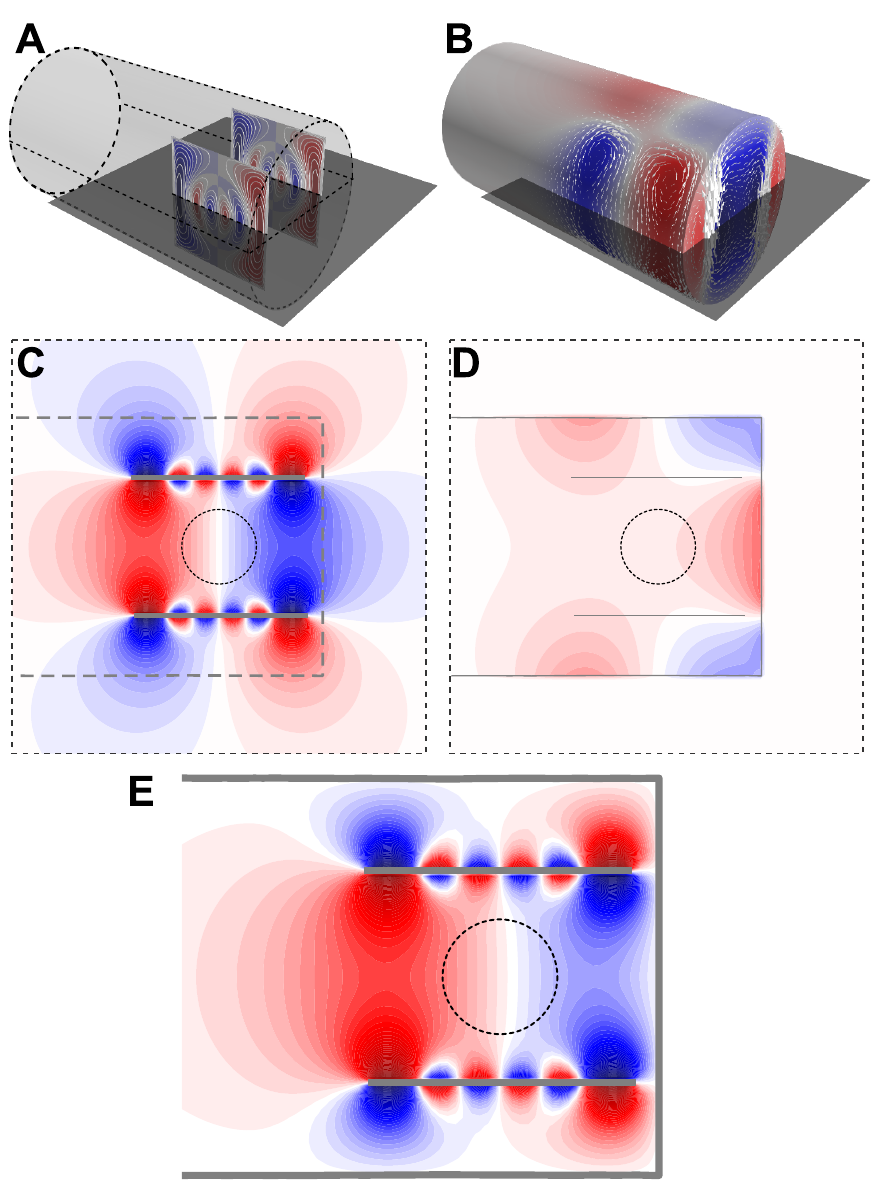}
    \caption{A current distribution (coil) on a bi-planar surface (\textbf{A}) inside a perfect cylindrical $\mu$-metal magnetic shield, an equivalent surface current on the shield surface (\textbf{B}), and associated magnetic scalar potentials (\textbf{C},\textbf{D},\textbf{E}) plotted on the horizontal plane shown in \textbf{A} and \textbf{B}. \textbf{A}: The stream function of the primary current distribution on the bi-planar surfaces inside the shield. \textbf{B}: The equivalent surface-current distribution (stream function and surface-current density) representing the induced field source on the $\mu$-metal shield. \textbf{C}: The primary magnetic scalar potential generated by the primary source in \textbf{A}. \textbf{D}: The magnetic scalar potential generated by the equivalent current in \textbf{B} inside the shield surface. \textbf{E}: The combined potential that satisfies the constant-potential boundary condition. The dashed circle represents the volume of interest.}
    \label{fig:perfect_mu_metal_shield}
\end{figure}

Figure~\ref{fig:perfect_mu_metal_shield} shows an example of a magnetically shielded configuration with a surface-current pattern on a bi-planar surface inside a perfect cylindrical magnetic shield. The current patterns on the planes are designed such that the field in the target volume, indicated by the dashed circle, is as homogeneous as possible. The secondary field due to the shield amplifies the magnetic field (the gradient of the potential) in the vicinity of the shield outside the target volume. The contribution from the cylinder cap also produces minor inhomogeneity in the magnetic field inside the target volume.

\section{Discussion and outlook}
\noindent We have introduced a set of tools for static and quasistatic modeling of divergence-free surface currents and their fields. The tools can be used for a variety of tasks from surface-coil design and equivalent-source modeling to eddy-current and thermal-noise calculations. This work has covered the central computations implemented in the software, the structure of which is described in the accompanying paper. \cite{zetter2020bfieldtools}

The computational and theoretical framework leverages the interpretation of stream functions as magnetic dipole densities normal to the surface. This analogy has been recognized previously \cite{peeren2003thesis, lopez2009equivalent}, but has not been fully exploited.
In this work, we exploit this interpretation further for the calculation of the magnetic scalar potential of a stream function, which we use, e.g., for visualizing the magnetic field. The scalar potential also enables the application of harmonic potential theory commonly applied in volume-conductor problems in the form of boundary-element methods (BEM). \cite{kybic2005common, stenroos2007matlab} The discretizations implemented in \texttt{bfieldtools} are directly applicable for BEM computations. Namely, the potential of a linearly varying dipole density can be used to calculate the \emph{double-layer} operator $D$ for linear (hat) basis functions, and the potential of a constant charge density can be used for the \emph{single-layer} operator $S$ with constant basis functions. Additionally, the mutual inductance operator $\boldsymbol{M}$ is equivalent to yet another operator called $N$, which maps a dipole density to the normal field component.  \cite{nedelec2001acoustic, kybic2005common}

In the future, it can be fruitful to exploit the analogy between quasistatic magnetic problems and electric volume conductor problems even further. The source of the field in the former can be interpreted as magnetic dipoles, whereas in the latter, the sources are current dipoles. The magnetic scalar potential is analogous to the electric potential (both satisfy Laplace's equation), and if we interpret the permeability $\mu$ as the counterpart of electrical conductivity $\sigma$ in a volume conductor, the magnetic field is perfectly analogous to the volume current density. The vector potential in the magnetic problem further corresponds to the magnetic field in a volume conductor problem. This means that the tools presented in this work could be
applied to solve the electric potential in a volume conductor, e.g., for modeling transcranial magnetic stimulation \cite{sanchez2016novel, koponen2017coil} or for solving the bioelectromagnetic forward problem. \cite{stenroos2007matlab}

The multipole and surface-harmonic expansions implemented in \texttt{bfieldtools} are also suitable for applications in a more general context, e.g., in biomagnetic experiments or geomagnetism. The multipole expansion has been applied to source modeling in bioelectromagnetism. \cite{geselowitz_bioelectric_1967, wikswo_scalar_1985, nolte_calculation_1997, jerbi_meg_2002} In magnetoencephalography, it has also been applied in signal space separation (SSS), \cite{taulu_presentation_2005} which can be used to design software spatial filters to reject external interference fields. In principle, the surface-harmonic expansion could be used for the same purpose with more general convergence properties.



\section*{Acknowledgments}
\noindent
This work has received funding from the Vilho, Yrjö and Kalle Väisälä Foundation (author AM), European Union's Horizon 2020 research and innovation programme under grant agreement No. 820393 (macQsimal), the European Research Council under ERC Grant Agreement no. 678578 (HRMEG), and the Swedish Cultural Foundation under grant no. 140635 (author RZ). 

\section*{Data availability}
Data sharing is not applicable to this article as no new data were created or analyzed in this study.

\section*{References}

\bibliographystyle{aipnum4-1}
\bibliography{refs.bib}

\appendix
\section{Stream function as magnetic dipole density}
\label{sec:dipole_appendix}
\noindent In this appendix, we demonstrate the equivalence of the current-density and dipole-density interpretations of the stream function $\psi$ based on the magnetic vector potential. Let us start from the vector potential of a surface-current density:
\begin{equation}
    \vec A(\vec r) = \frac{\mu_0}{4\pi} \int_S \frac{\vec j(\vec r\,')}{R} dS'
        = \frac{\mu_0}{4\pi} \int_S \frac{\nabla_{\|}' \psi(\vec{r}\,')\times \n'}{R} dS'\,.
        \label{eq:app_current}
\end{equation}
where $\vec R = \vec{r}-\vec{r}\,'$ and $R = |\vec R|$ to simplify the expression.
Using the product rule $\nabla_{\|}\,' (\psi(\vec{r}\,')/{R}) = \nabla_{\|}' \psi(\vec{r}\,')/{R} + \psi(\vec{r}\,') \nabla_{\|}'(1/R)$ on the tangent plane, we get
\begin{equation}
\begin{split}
    \vec A(\vec r) 
    = \frac{\mu_0}{4\pi}\bigg( &\int_S \nabla_{\|}' \frac{\psi(\vec{r}\,')}{R} \times \n' dS'\\ &+  \int_S \psi(\vec{r}\,')\n' \times \nabla_{\|}' \frac{1}{R} dS'\bigg)\,.
\end{split}    
\end{equation}
With Stokes's theorem on the surface, the first integral can be converted to a line integral of $\psi(\vec{r}\,')/R$ over the boundary of $S$. As discussed in Sec.~\ref{sec:bfield_surface_currents}, the stream function must be constant on the boundary. On a single boundary this constant can be set to zero and the line integral vanishes. When the surface contains holes, we get rid of the line integrals, by extending the constant values over the holes. This redefinition does not affect $\vec{j}$, but enables us to express the vector potential as
\begin{equation}
    \vec A(\vec r)  
     = \frac{\mu_0}{4\pi} \int_S \psi(\vec{r}\,')\n' \times \nabla' \frac{1}{R} dS'\,,
     \label{eq:app_dipole}
\end{equation}
where we have applied $\n \times \nabla_{\|}' = \n \times \nabla'$.
This is the vector potential of a magnetic dipole density $\vec{m}(\vec{r}\,') = \psi(\vec{r}\,')\n(\vec{r}\,')$.

The two forms of the mutual inductance $M_{k,l}$ in Eq.~\eqref{eq:inductive_energy} can be obtained using the equivalence of Eqs.~\eqref{eq:app_current} and \eqref{eq:app_dipole}. We start from
\begin{equation}
    M_{k,l} = \int_{S_k} \vec j_k(\vec r) \cdot \vec A_l(\vec r) dS\,,
\end{equation}
where $\vec{j}_k$ is one surface-current density and $\vec{A}_l$ the vector potential of surface-current density $\vec{j}_l$.
By substituting the vector potential in the dipole-density form Eq.~\eqref{eq:app_dipole}, we get a double integral, the integrand of which can be manipulated as
\begin{equation}
\begin{split}
    &\vec j_k(\vec r) \cdot \left(\frac{\mu_0}{4\pi} \vec{m}_l(\vec{r}\,') \times \nabla' \frac{1}{R}\right) \\ &= \vec{m}_l(\vec{r}\,')  \cdot \left(\frac{\mu_0}{4\pi} \vec{j}_k(\vec r) \times  \nabla \frac{1}{R}\right)\,.
\end{split}
\end{equation}
Identifying the expression in the parenthesis on the right as the integrand of the Biot--Savart law for $\vec{j}_k$ gives
\begin{equation}
    M_{k,l} = M_{l,k} = \int_{S_k} \vec m_k(\vec r) \cdot \vec B_l(\vec r) dS\,.
\end{equation}

\section{Integral formulas for triangles}\label{sec:Bfield_appendix}
Here, we present simplified derivations of the integral formulas that involve the solid angle $\Omega_f$ [Eq.~\eqref{eq:solid_angle}] and line-charge potentials $\gamma_i$ [Eq.~\eqref{eq:gamma}]. The derivations share some common aspects, which we would like to point out. The integrands are first manipulated so that a term that contains the solid angle can be separated. The rest of the integrand can be expressed as surface divergence on the triangle, for which Gauss's theorem can be applied, yielding expressions containing the line-charge potentials. Finally, the coefficients multiplying the analytical integrals are determined using the geometry of the problem.

\paragraph{Magnetic field of a constant current on a triangle}
\noindent Using a basic vector identity for the vector triple product, we can write the integrand in the Biot--Savart formula in Eq.~\eqref{eq:biot-savart} for $\nabla'_\| \psi(\vec{r}\,') \times \n_f = \vec{G}_{f,i} \times \n_f  =\vec{G}^\perp_{f,i}$ as
\begin{equation}
    \vec{G}^\perp_{f,i} \times \frac{\vec{R}}{R^3}
    = \left(\vec{G}_{f,i} \cdot\frac{\vec{R}}{R^3}\right) \n_f - \left(\n_f \cdot \frac{\vec{R}}{R^3}\right) \vec{G}_{f,i}\,,
\end{equation}
where $\vec R = \vec{r}-\vec{r}\,'$ and $R = |\vec R|$. We can further write the scalar part of the first term on the right-hand side as
\begin{equation}
    \vec{G}_{f,i} \cdot \frac{ \vec{R}}{R^3} = \vec{G}_{f,i} \cdot \nabla'\frac{1}{R} 
    = \nabla'_\| \cdot \left(\frac{\vec{G}_{f,i}}{R}\right)\,.
    \label{eq:divergence_triangle}
\end{equation}
Now, we can integrate the expression over the triangle $\Delta_f$. Using Gauss's theorem for the first term, and the definition of the solid angle for the second, we get
\begin{equation}
\begin{split}
    &\int_{\Delta_f}  \vec{G}^\perp_{f,i} \times \nabla' \frac{1}{R} dS' \\
     &= \n_f \int_{\partial \Delta_f}\frac{\vec{G}_{f,i}}{R} \cdot \left(-\n_f \times d\vec{l}\,'\right) + \vec{G}_{f,i}\Omega_f(\vec{r})\,,
\end{split}
\end{equation}
where $-\n_f \times d\vec{l}\,'$ is a line differential on the triangle boundary $\partial \Delta_f$ perpendicular to $d\vec{l}\,'$ pointing out of the triangle in the triangle plane. Rearranging the scalar triple product inside the first integral, we get
\begin{equation}
\begin{split}
    &\int_{\Delta_f}  (\vec{G}_{f,i} \times \n_f) \times \nabla' \frac{1}{R} dS' 
    \\ & = \n_f (-\vec{G}_{f,i} \times \n_f) \cdot \int_{\partial \Delta_f}\frac{1}{R} d\vec{l}\,' - \vec{G}_{f,i}\Omega_f(\vec{r})\,,
    \\ & = \Omega_f(\vec{r})\vec{G}_{f,i} -\sum_{l=i,j,k} c_{i,l} \gamma_l(\vec{r})\n_f\,,
\end{split}
\end{equation}
where $c_{i,l} = \vec{G}^\perp_{f,i} \cdot \vec{e_l} = (\vec{e_i} \cdot \vec{e_l})/(2A_f)$.


\paragraph{Potential of a uniform charge density on a triangle}
\noindent We decompose the displacement vector as $\vec{R} = \vec{p} + d_f\n_f$, where $\vec{p}$ is the component along the plane of the triangle and $d_f=\n_f \cdot \vec{R}$ is the signed distance from the triangle plane. This leads to the following identity
\begin{equation}
    \frac{1}{R} = -\nabla_\|' \cdot \frac{\vec{p}}{R} -  \frac{d_f^2}{R^3}\,.
\end{equation}
Integrating the expression over $\Delta_f$, we get
\begin{equation}
    \int_{\Delta_f} \frac{1}{R} dS = -\int_{\partial \Delta_f} \frac{\vec{p}\cdot (-\n_f' \times d\vec{l}\,')}{R} + d_f\,\Omega(\vec{r})\,,
\end{equation}
where the first term has been obtained by Gauss's theorem and the second by applying the definition of the solid angle. Integrating each triangle edge in $\partial \Delta_f$ separately and noting that the numerators of these integrals do not depend on the integration variable, we can express the line integral using the line-charge potentials: 
\begin{equation}
    \int_{\Delta_f} \frac{1}{R} dS = d_f(\vec{r})\Omega_f(\vec{r}) +  \sum_{l=i,j,k} 2 A_f x_l(\vec{r}) \gamma_l(\vec{r})\,,
\end{equation}
where $x_l = \vec{G}_{f,l} \cdot \vec{d}_{l+1}$ is the normalized signed distance measured in the triangle plane from the line defined by edge $\vec{e}_l$ towards node $l$ so that $x_l(\vec{r}_i)=1$.

\paragraph{Potential of a linear dipole density}
With identities $d_f = \n_f\cdot \vec{R}$ and $h_i(\vec{r}\,') = \vec{G}_{f,i} \cdot (\vec{r}' - \vec{r}_j) = \vec{G}_{f,i} \cdot (\vec{d}_j - \vec{R})$, we can write the potential of a linearly varying dipole density as
\begin{equation}
    \int_{\Delta_{f}} h_i(\vec r\,')\frac{\vec{R}}{R^3}\cdot d \vec S\,'
    =  d_f \int_{\Delta_{f}} \frac{\vec{G}_{f,i} \cdot (\vec{d}_j - \vec{R})}{R^3} dS\,'\,.
\end{equation}
Again, let us separate a term containing the solid angle:
\begin{equation}
\begin{split}
    &\int_{\Delta_{f}} h_i(\vec r\,')\frac{\vec{R}}{R^3}\cdot d \vec S\,' \\
    &= -x_i(\vec{r}) \Omega_f(\vec{r}) - d_f \int_{\Delta_{f}} \vec{G}_{f,i} \cdot \frac{ \vec{R}}{R^3} dS\,'.
\end{split}
\end{equation}
For the latter integral, we can utilize the identity in Eq.~\eqref{eq:divergence_triangle}:
\begin{equation}
\begin{split}
    &\int_{\Delta_{f}} h_i(\vec r\,')\frac{\vec{R}}{R^3}\cdot d \vec S\,' \\
    &= -x_i(\vec{r}) \Omega_f(\vec{r}) +  \sum_{l=i,j,k} c_{i,l}d_f(\vec{r}) \gamma_l(\vec{r})\,.
\end{split}
\end{equation}

\end{document}